\documentclass[10pt,conference]{IEEEtran}
\usepackage[T1]{fontenc}
\usepackage{amsmath}
\usepackage{url}
\usepackage[dvipsnames]{xcolor}
\usepackage{tikz}
\usepackage{pgfplots}
\pgfplotsset{compat=1.17}
\usetikzlibrary{arrows.meta,positioning,shapes.geometric,fit,backgrounds,calc}

\def\BibTeX{{\rm B\kern-.05em{\sc i\kern-.025em b}\kern-.08em T\kern-.1667em\lower.7ex\hbox{E}\kern-.125emX}}

\begin{document}

\title{SpBench: An Agent-Native Framework for Cross-Engine Lakehouse Benchmarking}

\author{
\IEEEauthorblockN{Lyndon Shi\textsuperscript{*}, Nic Crouch\textsuperscript{*}, Yi Hu\textsuperscript{*}, Zihao Ye\textsuperscript{*}, Choden Konigsmark,\\ Eric Vandenberg, Dionysios Logothetis, Kuba Niemiec, Krzysztof Gadomski, Yuxiong He}
\IEEEauthorblockA{Snowflake Inc.}
}

\maketitle
\begingroup
\renewcommand{\thefootnote}{*}
\footnotetext{These authors contributed equally to this work.}
\endgroup

\begin{abstract}
SpBench is an agent-native framework for cross-engine lakehouse benchmarking, in production use for benchmarking Snowpark, Snowpark Connect for Apache Spark, and other execution environments across standardized suites --- such as TPCx-BB and TPC-DI --- and custom benchmarks distilled from production workload patterns. It runs continuously across a broad portfolio of workloads, guiding the performance roadmap and detecting regressions for production-scale data engines. Benchmarking has historically been an expensive investment requiring continuous cross-functional collaboration and domain expertise, making it difficult to obtain actionable performance insights. SpBench provides (i) a typed, code-generated benchmark substrate that compiles platform-agnostic benchmark definitions into fair, comparable platform-specific implementations; (ii) a structured-telemetry contract recording every run as queryable fact and dimension data, letting LLM agents query results deterministically; and (iii) an agent-native harness of three composable skills to expedite building benchmarks and analyzing results. This agent-native architecture enabled the rapid development of more than 400 benchmark queries that run on a regular cadence to detect performance regressions; catching these has directly led to improvements in production.
\end{abstract}


\section{Introduction}

Modern cloud data platforms expose different execution engines for the same logical workload. On Snowflake this includes the Snowpark API for native execution \cite{snowpark} and Snowpark Connect for Apache Spark \cite{snowparkconnect}; SpBench additionally benchmarks Apache Spark \cite{rdd} across other lakehouse platforms. Comparing these engines fairly is a recurring industrial need that is expensive and time-consuming. Continuous cross-functional collaboration is required as different roles look for different signals from benchmarking, and benchmark implementations require domain experts.

Recent advances in foundation models \cite{foundationmodels} and LLM agents \cite{react} suggest a different operating model: an LLM agent can, in principle, translate a natural-language workload description into engine-equivalent implementations, drive their execution, and produce a structured comparison. Naive ``agent-in-a-shell'' approaches --- an LLM issuing raw shell commands without typed schemas or structured intermediaries --- often fail to produce comparable implementations across engines and struggle to fit multi-engine understanding in a single context. Robust primitives and a unified architecture are therefore required to guide AI toward rigorous implementations and reliable results.

This paper presents SpBench, an agent-native framework for cross-engine lakehouse benchmarking in production use. Three concepts enable LLM agents to act as first-class consumers of the benchmarking workflow:

\begin{enumerate}
\item \emph{Strictly typed benchmark definitions.} A benchmark is expressed step-by-step in a platform-agnostic way to achieve equivalent implementations across engines by construction.
\item \emph{Structured-telemetry contract.} Every run is recorded as queryable fact and dimension data in a standardized format, allowing deterministic queries for results.
\item \emph{Agent-native harness.} Composable skills form a closed loop for independent, self-reviewing agent iteration.
\end{enumerate}

\section{Background}
\label{sec:background}

Cross-engine benchmarking is fundamentally a triple-equivalence problem: a fair comparison between two engines must hold the data, the query semantics, and the compute tier all equal. Workload specifications such as TPC-DS \cite{tpcds} and TPCx-BB \cite{tpcxbb}\footnote{TPCx-BB is based on the BigBench benchmark \cite{bigbench}.} standardize what to run, but say nothing about how to compose engine-specific implementations of the same logical query, how to match compute tiers of comparable cost, or how to verify that two implementations have produced equivalent results. Those gaps are filled today by ad-hoc engineering effort.

Compounding these gaps is the need for cross-platform expertise; engineers are often only well-versed in their own platform. An agent-native framework must remove this bottleneck --- not by replacing equivalence policing with prompts, but by promoting equivalences from ``by convention'' to ``by construction''.

Benchmark management systems such as OLTP-Bench \cite{oltpbench} address portability across relational databases but target single-engine deployments with no agent-facing interface. Lakehouse-specific benchmarks such as LHBench \cite{lhbench} compare storage-format performance across engines but do not address agent-driven authoring or cross-engine query equivalence. SpBench is designed from the ground up for agents by providing strict, composable building blocks to target arbitrary execution engines from a common foundation to ensure comparable cross-engine results.

\section{Architecture}
\label{sec:architecture}

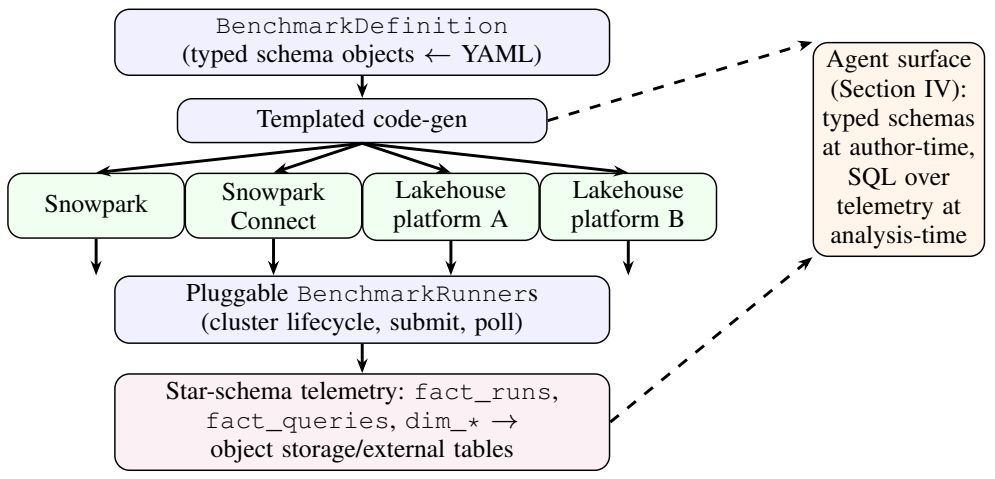
\begin{figure}[t]
\centering
\resizebox{0.74\linewidth}{!}{% Figure 1: SpBench substrate pipeline
\begin{tikzpicture}[
  font=\scriptsize,
  node distance=6pt,
  box/.style={draw, rounded corners, align=center, minimum height=0.42cm, inner sep=2pt, fill=blue!6},
  engbox/.style={draw, rounded corners, align=center, minimum height=0.62cm, inner sep=2pt, fill=green!6, text width=1.55cm},
  agent/.style={draw, rounded corners, align=center, minimum height=0.9cm, inner sep=2pt, fill=orange!8, text width=1.5cm},
  arr/.style={-{Stealth[length=4pt]}, thick}
]

\node[box, text width=4.6cm] (def) {\texttt{BenchmarkDefinition} (typed schema objects $\leftarrow$ YAML)};
\node[box, below=of def, text width=3.4cm] (codegen) {Templated code-gen};

\node[engbox, below=8pt of codegen, xshift=-2.55cm] (sp) {Snowpark};
\node[engbox, below=8pt of codegen, xshift=-0.85cm] (spc) {Snowpark Connect};
\node[engbox, below=8pt of codegen, xshift=0.85cm] (platA) {Lakehouse platform A};
\node[engbox, below=8pt of codegen, xshift=2.55cm] (platB) {Lakehouse platform B};

\node[box, below=10pt of spc, xshift=0.85cm, text width=4.6cm] (runners) {Pluggable \texttt{BenchmarkRunner}s (cluster lifecycle, submit, poll)};

\node[box, below=8pt of runners, text width=4.6cm, fill=purple!6] (tel) {Star-schema telemetry: \texttt{fact\_runs}, \texttt{fact\_queries}, \texttt{dim\_*} $\to$ object storage/external tables};

\node[agent, right=26pt of platB, yshift=0.55cm] (agentbox) {Agent surface (Section IV): typed schemas at author-time, SQL over telemetry at analysis-time};

\draw[arr] (def) -- (codegen);
\draw[arr] (codegen.south) -- (sp.north);
\draw[arr] (codegen.south) -- (spc.north);
\draw[arr] (codegen.south) -- (platA.north);
\draw[arr] (codegen.south) -- (platB.north);
\draw[arr] (sp.south) -- (runners.north -| sp.south);
\draw[arr] (spc.south) -- (runners.north -| spc.south);
\draw[arr] (platA.south) -- (runners.north -| platA.south);
\draw[arr] (platB.south) -- (runners.north -| platB.south);
\draw[arr] (runners) -- (tel);
\draw[arr, dashed] (codegen.east) -- (agentbox.north west);
\draw[arr, dashed] (tel.east) -- (agentbox.south west);

\end{tikzpicture}}
\caption{The SpBench substrate. A typed \texttt{BenchmarkDefinition} compiles via templated code generation into per-engine artifacts executed by pluggable runners, emitting star-schema telemetry (\texttt{fact\_runs}, \texttt{fact\_queries}, \texttt{dim\_*}) that the agent surface (Section~\ref{sec:agentic}) reads at author- and analysis-time.}
\label{fig:substrate}
\end{figure}

SpBench's substrate is a typed compile-and-execute pipeline: one benchmark definition compiles into engine-specific implementations, runs against pluggable engine runners, and emits structured telemetry to a shared analytics layer. Every layer is deterministic --- agents author against typed schemas, not free-form code, and outputs are queryable rather than merely parseable. Fig.~\ref{fig:substrate} illustrates the pipeline.

\begin{figure}[t]
\centering
\resizebox{0.62\linewidth}{!}{% Figure: BenchmarkDefinition schema hierarchy (2x2 snake layout)
\begin{tikzpicture}[
  font=\scriptsize,
  node distance=8pt,
  lvl/.style={draw, rounded corners, align=left, inner sep=5pt, text width=4.0cm},
  arr/.style={-{Stealth[length=4pt]}, thick}
]

\node[lvl, fill=blue!8] (bd) {
  \textbf{\texttt{BenchmarkDefinition}}\\
  name: str\\
  created\_resources: list[CreatedResource]\\
  query\_groups: list[\texttt{QueryGroup}]
};

\node[lvl, fill=cyan!8, right=34pt of bd] (qg) {
  \textbf{\texttt{QueryGroup}}\\
  name: str,\ serial: bool\\
  queries: list[\texttt{Query}]
};

\node[lvl, fill=green!8, below=18pt of qg] (q) {
  \textbf{\texttt{Query}}\\
  name: str,\ group: str\\
  steps: list[\texttt{QueryStep}]
};

\node[lvl, fill=yellow!15, below=18pt of bd] (qs) {
  \textbf{\texttt{QueryStep}}\\
  file: Path,\ function: str\\
  arguments: dict[str, Any]\\
  sql\_file: optional Path
};

\draw[arr] (bd) -- node[above, font=\tiny] {1-to-many} (qg);
\draw[arr] (qg) -- node[right, font=\tiny] {1-to-many} (q);
\draw[arr] (q) -- node[below, font=\tiny] {1-to-many} (qs);

\end{tikzpicture}}
\caption{The \texttt{BenchmarkDefinition} schema hierarchy: each level holds a typed, one-to-many list of the level below, from the top-level benchmark down to individual \texttt{QueryStep}s.}
\label{fig:schema}
\end{figure}
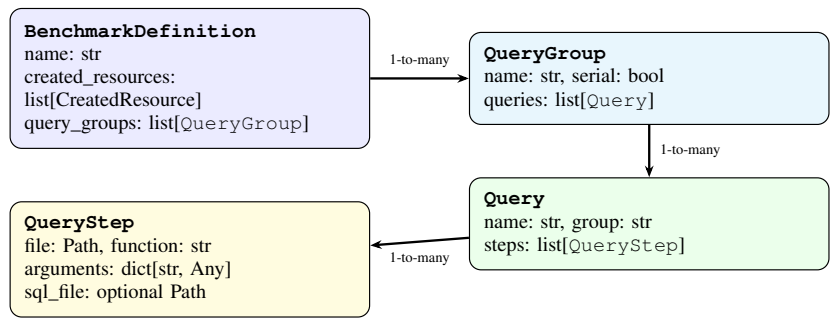

A benchmark decomposes into a hierarchy of typed constructs: \texttt{BenchmarkDefinition}, \texttt{QueryGroup}, \texttt{Query}, and \texttt{QueryStep}, each with a typed schema. Fig.~\ref{fig:schema} shows the decomposition: a \texttt{BenchmarkDefinition} holds one-to-many \texttt{QueryGroup}s, each holding one-to-many \texttt{Query} instances, each holding one-to-many \texttt{QueryStep}s --- the same steps whose per-step timing is recorded at run time, as described below. From a single benchmark definition, the substrate generates engine-specific notebooks or scripts via templated code generation, enabling execution against each target engine. The same \texttt{BenchmarkDefinition} produces all artifacts, so the query equivalence of Section~\ref{sec:background} is enforced by construction rather than by convention.

Generated artifacts run inside corresponding native environments, each via a pluggable \texttt{BenchmarkRunner} (Fig.~\ref{fig:substrate}): Snowpark workloads in the Snowpark Execution Environment \cite{seepp}, a secure sandboxed runtime co-located with Snowflake's elastic compute architecture \cite{sigmod}; PySpark workloads through Snowpark Connect; and Apache Spark workloads on other lakehouse platforms.

Every run emits a star-schema telemetry record to object storage: a \texttt{fact\_runs} row identifying the run, one \texttt{fact\_queries} row per query with timing and structured error fields, and content-addressed (hash-keyed) \texttt{dim\_*} records for the deployment, compute tier, and environment, exposed as external tables so deterministic SQL queries power both dashboards and the agent skills of Section~\ref{sec:agentic}. Beyond the per-query \texttt{fact\_queries} record, a per-step timing table captures sub-query timing, letting the agent localize whether latency is in data loading, compilation, or execution. Users can also add extensions to collect platform-specific telemetry beyond this default contract after each benchmark query run.

\section{Agentic Benchmarking}
\label{sec:agentic}

\begin{figure}[t]
\centering
\resizebox{0.70\linewidth}{!}{% Figure 3: Two governed harness loops
\begin{tikzpicture}[
  font=\scriptsize,
  box/.style={draw, rounded corners, align=center, minimum height=0.5cm, inner sep=2pt, text width=1.62cm},
  gate/.style={draw, rounded corners, align=center, minimum height=0.55cm, inner sep=3pt, fill=red!8, text width=4.9cm},
  arr/.style={-{Stealth[length=4pt]}, thick}
]

% Operational loop (top row)
\node[box, fill=blue!6] (reg) {Regression signal (telemetry)};
\node[box, right=8pt of reg, fill=blue!6] (repro) {Reproduce and localize (Build + Analyze)};
\node[box, right=8pt of repro, fill=blue!6] (classify) {Classify: bench\\bug/config/\\product bug};

% Development loop (bottom row)
\node[box, below=20pt of reg, fill=green!6] (req) {External request};
\node[box, right=8pt of req, fill=green!6] (distill) {Distill $\to$ spec};
\node[box, right=8pt of distill, fill=green!6] (build) {Build $\to$ code + run};
\node[box, right=8pt of build, fill=green!6] (analyze) {Analyze $\to$ results summary};

\node[gate, below=18pt of build] (gate) {Staged acceptance-contract gate};

\draw[arr] (reg) -- (repro);
\draw[arr] (repro) -- (classify);
\draw[arr] (req) -- (distill);
\draw[arr] (distill) -- (build);
\draw[arr] (build) -- (analyze);

\draw[arr] (classify.south) to[out=-90,in=90] (gate.north -| classify.south);
\draw[arr] (analyze.south) -- (gate.north -| analyze.south);

\draw[arr, dashed, red!45!black] ($(gate.north)+(0.9,0)$) -- ($(build.south)+(0.9,0)$)
  node[midway, right=1pt, font=\tiny, text=red!45!black] {retry};

\coordinate (gapcol) at ($(distill.east)!0.5!(build.west)$);
\draw[arr, dashed, red!45!black] (gapcol |- gate.north) |- (repro.south)
  node[pos=0.15, right=1pt, font=\tiny, text=red!45!black] {retry};

\node[above=1pt of reg, font=\scriptsize\bfseries] {Operational loop};
\node[above=1pt of req, font=\scriptsize\bfseries] {Development loop};

\end{tikzpicture}}
\caption{SpBench's two governed harness loops. The operational loop reproduces, localizes, and classifies a telemetry-detected regression; the development loop turns an external request into a spec, code, and a results summary. Both loops are gated by staged acceptance contracts before completion.}
\label{fig:loops}
\end{figure}
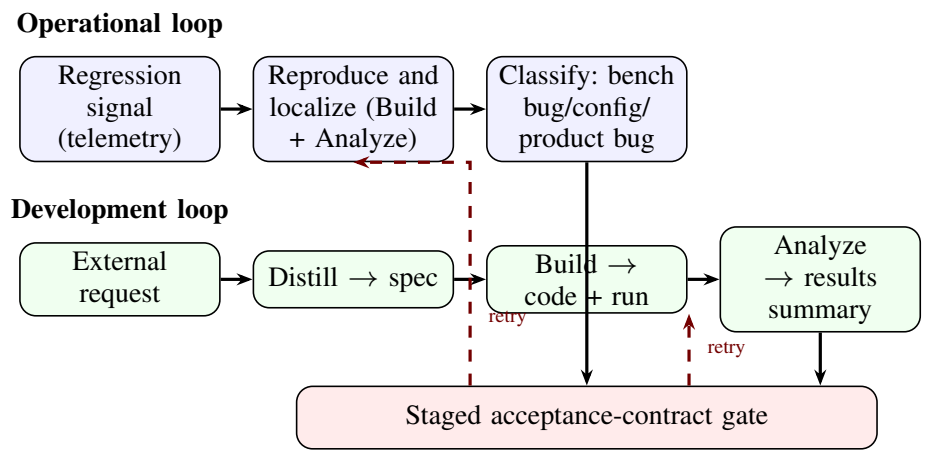

The substrate of Section~\ref{sec:architecture} is the deterministic surface for an agent layer comprising three composable benchmark skills and two governed harness loops that drive the skills against operational and development signals (Fig.~\ref{fig:loops}).

\textbf{Benchmark Skills.} Each top-level skill encodes a named, composable workflow that an LLM agent can invoke from an IDE or CLI through typed tool interfaces implemented over the Model Context Protocol \cite{mcp}, in the spirit of tool-augmented language models \cite{toolformer}: a dedicated MCP server exposes run monitoring, SQL query access, and artifact retrieval as typed tools, grounding agent interactions in structured calls rather than shell commands. Each skill routes to engine-specific sub-skills, with Analyze further decomposing into run-data retrieval, comparison reporting, and failure diagnosis. The three skills each serve a distinct phase of the workflow:

\begin{itemize}
\item \emph{Distill} ingests engine performance telemetry --- aggregate workload patterns, execution statistics, job-level metrics --- and produces a sanitized benchmark spec (tables, engines, success criteria) containing no customer query content, table values, or user-identifying information.
\item \emph{Build} transforms that spec into runnable artifacts --- per-engine query implementations, configurations, orchestration wiring --- using SpBench constructs.
\item \emph{Analyze} queries the benchmark telemetry tables after a run completes and emits a plain-language summary covering pass/fail rates, latency deltas, and outliers.
\end{itemize}

The three skills compose by data flow: Distill's output is Build's input, Build's output is the run, and Analyze reads its telemetry.

\textbf{Harness Loops.} The skills are driven by two governed loops against complementary signals (Fig.~\ref{fig:loops}): an operational loop that reproduces, localizes, and classifies a telemetry-detected regression via Build and Analyze, and a development loop that turns an external request into a spec, generated code, and a results summary via Distill, Build, and Analyze --- both gated by staged acceptance contracts, automated checks that a stage's output must satisfy before the loop advances: for Build, engine-specific code checks and matching configurations across engines before a run is launched. A rejected stage is sent back to Build --- in the operational loop, via Reproduce and localize --- for another iteration.

Both signal-generating skills emit structured, directly actionable recommendations rather than free-text findings, enabling an autonomous loop. Distill's eligibility framework screens each candidate query pattern against a defined eligibility rule set --- rejecting trivial or already-covered patterns and prioritizing signals such as window functions, multi-join plans, and novel UDF combinations --- then writes a filled benchmark-request file that Build consumes directly, with no manual transcription step. Analyze's regression scan carries engine-specific diagnostic context, enabling immediate discovery of benchmark implementation issues. Benchmark-side defects route back into Build's iteration path to patch the affected query and re-run.

\textbf{Decompose, Don't Monolith.} A natural design choice when applying LLM agents to a multi-step workflow is one agent, equipped with all required tools, that plans, writes, debugs, and analyzes in one continuous reasoning chain. This monolithic approach breaks down on the Build skill, which must simultaneously reason about benchmark semantics, generate per-engine query implementations, audit code against a multi-rule constraint set, and coordinate validation across engines. The monolithic approach suffers heavily from context overload and quickly derails as iterations accumulate.

We developed the Build skill using TeraSort \cite{terasort} as a golden end-to-end test case, targeting reliable execution across every supported engine, execution mode, and compute configuration. Starting from a structured user-request template (tables, engines, success criteria), we aimed for a single skill invocation that completed the full authoring cycle. Iteration proceeded through acceptance gates at each delegation boundary, per-engine implementation guidance, and graduated sub-agent delegation, until a single invocation reliably reached the validated state without manual intervention.

\begin{figure}[t]
\centering
\resizebox{0.80\linewidth}{!}{% Figure 4: Decomposed Build skill architecture
\begin{tikzpicture}[
  font=\tiny,
  stage/.style={draw, rounded corners, align=center, minimum height=0.42cm, inner sep=2pt, text width=1.5cm, fill=blue!6},
  sub/.style={draw, rounded corners, align=center, minimum height=0.34cm, inner sep=1.3pt, text width=0.72cm, fill=orange!8, font=\fontsize{7.3}{8}\selectfont},
  arr/.style={-{Stealth[length=3.5pt]}, thick}
]

\node[stage] (s1) {(i) Static codegen\\ \scriptsize per-engine sub-agent, parallel};
\node[stage, right=6pt of s1] (s2) {(ii) Code review\\ \scriptsize per-query sub-agent, parallel};
\node[stage, right=6pt of s2] (s3) {(iii) SF1 validation\\ \scriptsize per-engine sub-skill};
\node[stage, right=6pt of s3] (s4) {(iv) Full-scale run\\ \scriptsize per-engine sub-skill};

\node[sub, below=6pt of s1, xshift=-0.44cm] (s1a) {eng A};
\node[sub, below=6pt of s1, xshift=0.44cm] (s1b) {eng B};

\node[sub, below=6pt of s2, xshift=-0.44cm] (s2a) {dir A};
\node[sub, below=6pt of s2, xshift=0.44cm] (s2b) {dir B};

\node[sub, below=6pt of s3, xshift=-0.44cm] (s3a) {eng A};
\node[sub, below=6pt of s3, xshift=0.44cm] (s3b) {eng B};

\node[sub, below=6pt of s4, xshift=-0.44cm] (s4a) {eng A};
\node[sub, below=6pt of s4, xshift=0.44cm] (s4b) {eng B};

\draw[arr] (s1) -- (s2);
\draw[arr] (s2) -- (s3);
\draw[arr] (s3) -- (s4);

\draw[-{Stealth[length=3pt]}] (s1.south) -- (s1a.north);
\draw[-{Stealth[length=3pt]}] (s1.south) -- (s1b.north);
\draw[-{Stealth[length=3pt]}] (s2.south) -- (s2a.north);
\draw[-{Stealth[length=3pt]}] (s2.south) -- (s2b.north);
\draw[-{Stealth[length=3pt]}] (s3.south) -- (s3a.north);
\draw[-{Stealth[length=3pt]}] (s3.south) -- (s3b.north);
\draw[-{Stealth[length=3pt]}] (s4.south) -- (s4a.north);
\draw[-{Stealth[length=3pt]}] (s4.south) -- (s4b.north);

\end{tikzpicture}}
\caption{Decomposed Build skill architecture. Authoring is delegated at four points --- static codegen, code review, SF1 validation, full-scale run --- each handled by narrow-scope sub-agents (\emph{eng}: per-engine; \emph{dir}: per-query-directory).}
\label{fig:decompose}
\end{figure}
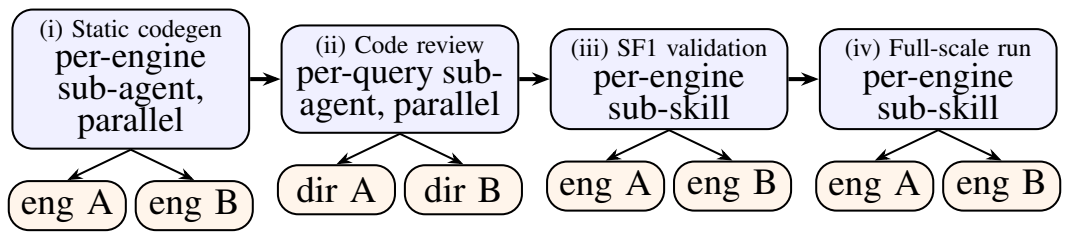

The resulting architecture (Fig.~\ref{fig:decompose}) decomposes authoring at four stages: (i) \emph{static code generation}, parallel per-engine query writers; (ii) \emph{code review}, one sub-agent per query directory auditing all engine implementations in parallel; (iii) \emph{scale-factor-1 (SF1) validation}, engine-specific runner sub-skills operating independently; and (iv) \emph{full-scale execution} by the same sub-skills. Decomposition localizes context and recovery; errors can be addressed in a targeted manner, leaving other agents' state intact.

This points to a general pattern for multi-step LLM workflows: draw sub-agent boundaries at natural failure points, so a failure can be isolated and retried without restarting the whole workflow.

\section{Outcomes}
\label{sec:outcomes}

SpBench currently comprises more than 400 benchmark queries, drawn from standardized suites such as TPC-DI \cite{tpcdi} and TPCx-BB, and from custom benchmarks distilled from observed production patterns. These benchmarks run continuously to guide the Snowpark and Snowpark Connect performance roadmap and uncover regressions. The two engines together serve hundreds of millions of jobs per day in production \cite{snowpark}, making rigorous performance tracking a direct driver of quality at scale. Concrete areas of investment driven by benchmark insights include file-to-table data transformations and UDF optimizations.

With the benchmark authoring cycle reduced from weeks or months to just one AI development loop, we have been able to rapidly enhance our production benchmark portfolio with custom benchmarks derived from observed real-world usage, filling in blind spots left by industry-standard benchmarks \cite{workloadinsights}.

\section{Conclusion}
\label{sec:conclusion}

SpBench is an agent-native framework for cross-engine lakehouse benchmarking, pairing a deterministic substrate with three composable skills. Reliable one-shot benchmark authoring has been made possible by decomposed sub-agent harnesses with staged acceptance contracts. This system is used in production today by engineers across Snowpark to build and analyze benchmarks, and the resulting insights have spawned numerous development efforts for real product improvements. These outcomes underscore a broader principle: an LLM agent's productivity depends heavily on the surrounding structure of typed schemas, unified telemetry, governed loops, and decomposed skills. These findings reflect a single-organization deployment across a fixed engine set; broadening the evaluation across organizations and engines is ongoing work, moving benchmarking toward a model in which a customer-reported signal can be resolved by anyone, guided by AI.

\section*{Acknowledgment}

The authors thank colleagues at Snowflake for design discussions, code reviews, and engineering support throughout SpBench's development --- in particular the engineering and product teams whose products SpBench composes, the Snowflake Engineering Systems teams building foundational internal infrastructure powering benchmarking, and the early users whose feedback shaped the agentic interfaces and telemetry contracts presented here.

\end{document}